# Dependence of rate on complex GB migration by ramped-ECO

Tingting Yu[1]

*School of Mechanical Engineering, Changzhou Institute of Technology, Changzhou 213032, Jiangsu, China*

**Abstract**

GB migration plays a central role in microstructural evolution. Many experiments and simulations have been conducted to clarify the relationships between GB velocity and various parameters to tailor GB networks. However, the complexity of GB migration has surpassed initial expectations. In this study, the ramped Energy Conserving Orientational (r-ECO) Driving Force (DF) in Molecular Dynamics (MD) simulations was utilized to investigate grain boundary (GB) motion for $\Sigma3(110)$, $\Sigma15(211)$, and $\Sigma11(311)$. My findings indicate that the rate of the driving force determines the velocity of GB migration. Furthermore, a reverse shear coupling behavior during GB migration was observed when the rate was decreased in $\Sigma15(211)$. In addition to the change in the direction of shear coupling, a linear relationship between the rate and the transition point during shear coupling migration was discovered. Specifically, a larger rate leads to a forward shift in the transition point. Moreover, GB transition from coupled to only normal migration states in the presence of dislocations nucleated in the GB was observed. These findings contribute to a deeper understanding of microstructural evolution and have implications for designing materials with enhanced properties.

[1]Corresponding author
E-mail address: yutt@czust.edu.cn

# 1. Introduction

Over the past decades, various methods have been adopted to investigate the impact of grain boundaries (GBs) on the properties of polycrystals and how to tailor these properties based on our understanding of GB migration. Grain boundary-mediated plasticity has become a well-established phenomenon, with extensive research conducted on GB mobility[1–3], GB shear coupling[4], GB sliding[2], and GB transformation[3].

GB migration plays a vital role in GB plasticity as a deformation mechanism, particularly in the absence of dislocations. There is a difference in GB velocity for kinds of GBs or loading environments[5], such as GB misorientation[6], inclination, curvature, metal type, interatomic potential, system size, driving force, and variation of atomic grain boundary structure and loading history[5]. Deng et al. [7]demonstrated the dependence of GB velocity on the rate and size using the synthetic driving force method (SDF) proposed by Janssens et al. [8], with a primary focus on the velocity normal to the GB plane. The rate dependence on shear coupling deserves more attention. Shear coupling can be described by the ratio between the lateral and normal GB velocities ($v_{//}$ and $v_\perp$, respectively), denoted as $\beta = v_{//}/v_\perp$, where $\beta$ represents the shear coupling factor. Fang et al. [4] uncover two distinct migration patterns showing opposite shear coupling factors that can be activated by different stress/strain states for mixed grain boundaries.

The mechanisms underlying temperature dependence [9], loading direction [1,10], and rate [7,9,11] in terms of deformation or shear coupling behaviors are particularly important in the deformation behavior of nanocrystalline materials have been adequately considered. Here, I seek to further understand how the GB migration is affected by rate, and try to uncover some mechanisms behind the transition in shear coupling of GB.

In this work, the ramped energy-conserving orientational driving force method (r-ECO) based on ECO proposed by Ulomek et al. [12] was applied to study the rate dependence of GB migration, which includes the normal velocity and shear coupling. Through the r-ECO, the influence of the driving force on GB migration can be researched in one simulation. In this work, I will first detail how normal velocity and shear coupling behaviors depend on the applied external energy load rate. Furthermore, I will illustrate the reasons for the transition in shear coupling patterns. Finally, I will present the conclusions drawn from this study.

**2. Simulation Methods**

Molecular dynamics (MD) simulations were performed using the Large-scale Atomic/Molecular Massively Parallel Simulator (LAMMPS) [13]. The interatomic interactions were described using an Embedded Atom Method (EAM) potential developed by Foiles-Hoyt et al.[14]. Several grain boundaries (GBs) constructed by Olmsted et al. [15] were simulated, including Ni bicrystals Σ3 twist (110) (p5), Σ15 tilt (211)(p14), and Σ11 tilt (311)(p33).

As shown in Fig. 1, the GBs are located in the y-z plane. In the study, the simulation boxes for Σ3(110), Σ15(211), and Σ11(311) contained 14,875 atoms, 20,096 atoms, and 18,304 atoms, respectively. The dimensions of the simulation cells are listed in the

table1. In the simulation cells, the x direction used shrink-wrapped boundary conditions, while periodic boundary conditions were applied in the y and z directions. The GB models were relaxed at the test temperature ranging from 100K to 500K in the canonical ensemble (NVT) for 10ps and the isothermal-isobaric ensemble (NPT) for 15ps before applying the driving force. The time step used in the simulations was 5fs and timesteps influence the GB migration. All simulations were conducted using the r-ECO method. Ramped driving forces were employed, linearly increasing from 0 over time, to drive the GB migration [7,16]. The incremental rate of energy change can be calculated as $\dot{E} = \frac{\delta E}{\delta t}$ [7], so the driving force $\Delta E = \frac{\delta E}{\delta t} \cdot t$, as shown in Fig2. $\delta t$ is not the timestep of integration for molecular dynamics. To compute the normal GB displacement, the overall change of added artificial energy was tracked [17]. The lateral displacement of GB is computed by relative displacement of two ends of bicrystal. In this manuscript, ΔGB is the normal displacement of GB in the x-direction, Δz is the lateral displacement of GB in the z-direction.

Table 1 the dimensions of the simulation cells

| GB | PID | lx | ly | lz |
| --- | --- | --- | --- | --- |
| Σ3(110) | 5 | 17.6nm | 3nm | 3nm |
| Σ15(211) | 14 | 18nm | 3.1nm | 3.8nm |
| Σ11(311) | 33 | 18.6nm | 3.2nm | 3.3nm |

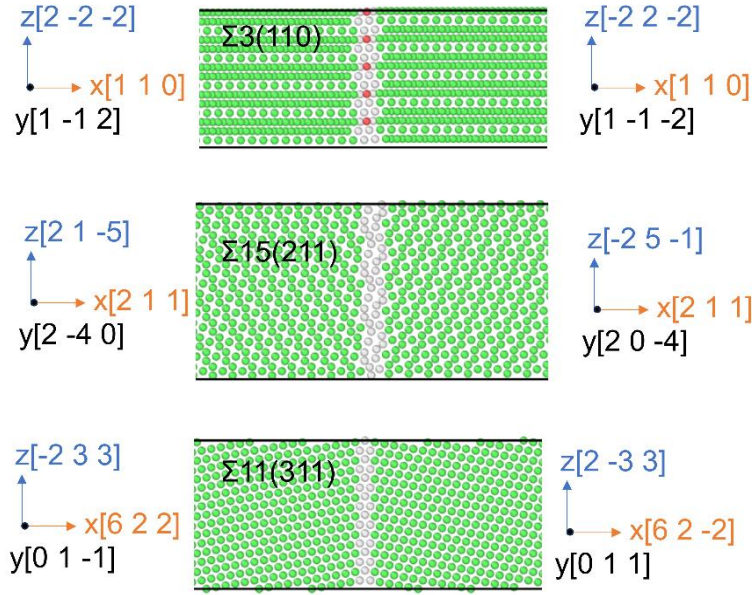

Fig. 1. The atomistic configurations of the Σ3(110)), Σ15(211) and Σ11(311)

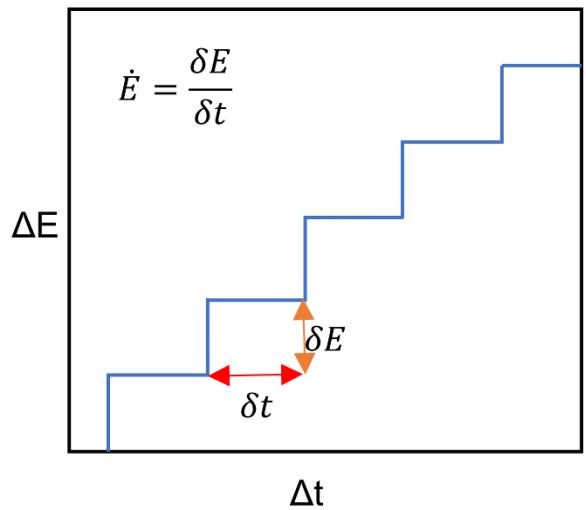

Fig. 2. The schematic of the driving force and the rate calculation

### 3. Results

*3.1 Rate dependence of GB velocity*

To investigate the rate dependence of GB normal velocity, simulations were performed on Σ3(110), Σ15(211), and Σ11(311) GBs using the r-ECO method at 100 K with different rates. The simulations were conducted with variations in driving force increment, time increment, and the combination of driving force increment and time increment. Fig. 3 illustrates the relationship between GB displacement and the applied driving force at varying rates ranging from 0.00004 eV/ps to 0.01 eV/ps. In Fig. 3a – Fig. 3c, different driving forces increment δE (0.001 eV, 0.0001 eV, and 0.00001 eV) were applied with the same time increment (δt = 0.05ps). Fig. 3d and Fig. 3e show simulations with the same driving force increment $\delta E$ but different time increment $\delta t$. Finally, Fig. 3f presents simulations with different combinations of δE and δt, but at the same rate. The basic idea behind the r-ECO method is that the driving force continuously increases in a step-wise manner ($\Delta E = \frac{\delta E}{\delta t} \cdot t$). Fig3 presents functions of the normal displacement(ΔGB) with time, and the relative velocity can be deduced from the slope of the lines. Fig3 shows a significant difference in the slopes of the curves corresponding to different rates at 100 K using the r-ECO method. Combining these observations with Fig. 3f, it can be concluded that the rate is the determining factor for normal velocity.

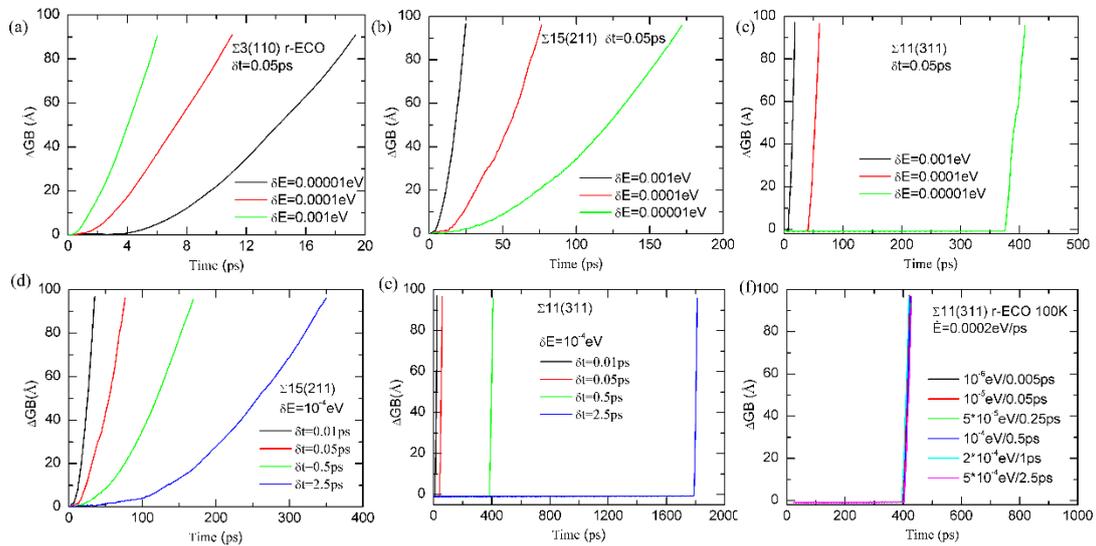

Fig. 3. (a)-(c) The normal displacement of the Σ3(110), Σ15(210), Σ11(311) GB in Ni vs. the time at the driving force rate ranging from 0.0002eV/ps - 0.02eV/ps, respectively. (d)-(e) The normal displacement of the Σ15(210), Σ11(311) GB in Ni vs.

the driving force at the rate ranging from 0.00004eV/ps - 0.01eV/ps, respectively. (f) The normal displacement of Σ11(311) GB in Ni under the same rate at 100K.

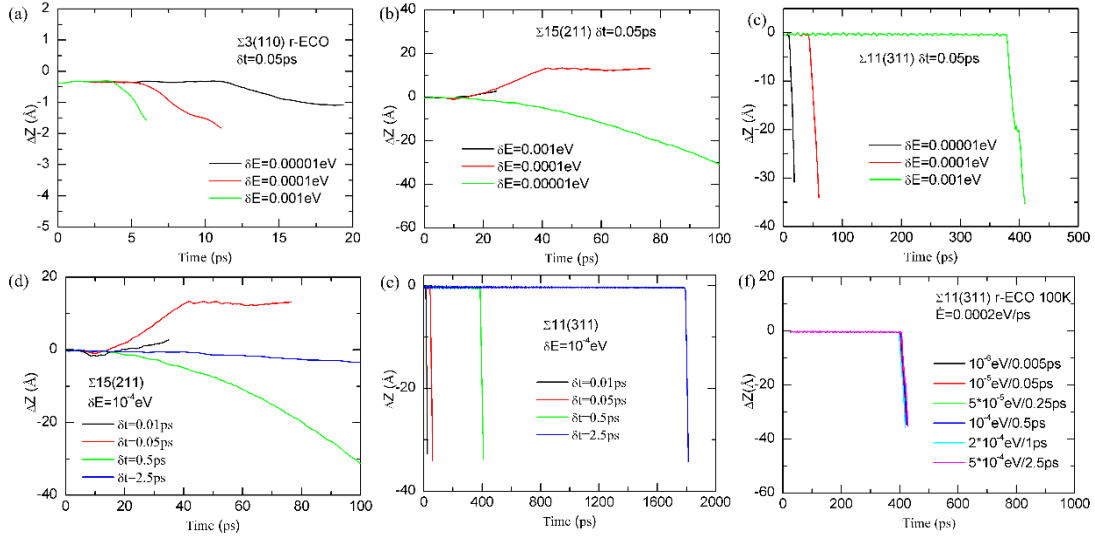

Fig.4. (a)-(e) Rate dependence of the lateral displacement of GB ~~shear coupling~~ for Σ3(110), Σ15(210), Σ11(311) at 100K by r-ECO. (f) ~~The translation in the z direction~~ The lateral displacement of Σ11(311) GB in Ni under the same rate at 100K.

Fig. 4 presents the relationship between ~~GB translation in the z-direction~~ the lateral displacement of GB (Δz) and the time at varying rates. Specifically, it examines the behavior of Σ3(110), Σ15(211), and Σ11(311) GBs under different rate. For Σ3(110) and Σ11(311) GBs in Fig.4a and Fig.4c, there is a noticeable transition where GB start to migrate abruptly. With increasing the rate, the driving force of the transition point decreases, and the velocity in z of GB increases. For Σ15(211) GB at rates of 0.02 eV/ps and 0.01 eV/ps, there are no abrupt transitions observed, but it presents the same tendency of velocity vs. rate as Σ3(110), Σ15(211). Furthermore, even with different δE and δt, the same rate leads to the same migration behavior for Σ11(311) GB, as shown in Fig. 4f, which means the rate decides the migration behavior. In Fig. 4, it can be observed that a higher rate increases GB migration velocity in z and changes the shear coupling mode. This is evident in the case of Σ15(211) GB migration, where a rate of 0.02 eV/ps (δE=0.001eV) leads to the transition from shear-coupled migration to only normal migration and an inverse z-direction of shear coupling migration compared to other rates. I will detail the dependence of the shear coupling mode on the rate in the next part.

*3.2 Dependence of rate on GB complex shear coupling mode for Σ15(211)*

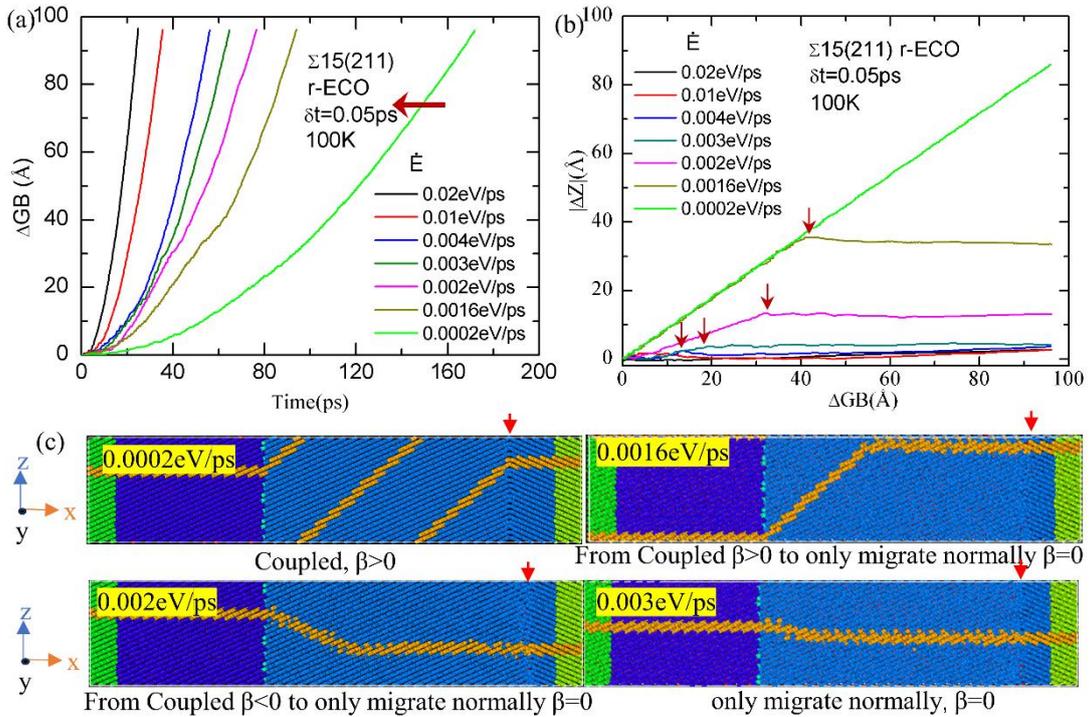

Fig.5. (a)The normal displacements of Σ15(211) GB at kinds of rates at 100K by r-ECO. (b) Shear coupling behaviors during Σ15(211) GB migration at 100K by r-ECO. (c)Snapshots of Σ15(211) GB at rates ($\dot{E}$) of 0.0002 eV/ps, 0.0016 eV/ps, 0.002 eV/ps, and 0.003 eV/ps at 100K, respectively.

Fig. 5 shows the normal displacements and shear coupling behaviors during migration of Σ15(211) GB at kinds of rates at 100K by r-ECO. Fig. 5a shows normal displacements during Σ15(211) GB migration at kinds of rates at 100K by r-ECO. Fig. 5b and Fig. 5c show shear coupling behaviors during Σ15(211) GB migration at 100K by r-ECO. The timestep is 0.005ps, $\delta T$=0.05ps, so the rates are 0.02eV/ps, 0.01eV/ps, 0.004eV/ps, 0.003eV/ps, 0.002eV/ps, 0.0016eV/ps and 0.0002eV/ps. In Fig. 5c, there is an obvious trend that a higher rate leads to weaker shear coupling, from GB coupled migration($\dot{E}$=0.0002eV/ps) to only normal migration ($\dot{E}$=0.003eV/ps). The shear-coupling factor changes from plus, minus, to zero.

It has been previously observed that a lack of change in shear coupling behavior during GB migration indicates a single disconnection mode[18]. In Fig. 5c, it is shown that a single disconnection mode is activated at a rate of 0.0002 eV/ps. However, when the rate is increased to 0.0016 eV/ps, multiple disconnection modes are activated. Interestingly, at a rate of 0.002 eV/ps, shear coupling occurs in the inverse direction. As the rate increases, the GB only migrates normally, and the transition point moves closer to the GB's original position.

With the increasing rate, the direction of shear coupling behaviors changes, but the absolute value of the shear coupling factor (the slope of the lines in Fig. 5b) decreases accordingly. Additionally, in some migrations, there is a regular difference in the direction of shear coupling, indicating that the same disconnection modes may have an inverse Burgers vector, leading to random changes in direction.

In Fig. 6, there are three stages during GB migration, I, II, and III. In stages I and III, there is no dislocation, but in stage II, there are eight 'other' dislocations and 1/2<1 1 0> dislocations, which depressed the GB migration in the z-direction, facilitating the only normal migration of GB. During the transition of the shear coupling mode, GB released dislocations and then absorbed dislocations, which led to a change in the shear coupling mode. In Fig6, there are also three stages in GB migration of Σ15(211) with $\delta E$ =0.00001 eV, $\delta T$=0.05ps. At stage Ⅰ, there is no dislocation in GB; At stage Ⅱ, there are some kinds of dislocations, including 1/2<110>(perfect) dislocation and other dislocations which Ovito can not describe. At stage Ⅲ, dislocations disappear again. Dislocations appear only at stage Ⅱ. At stage Ⅱ, the GB migration from shear-coupling migration to no shear-coupling migration, so this indicates that dislocations especially 1/2<110> lead to only normal migration during GB migration.

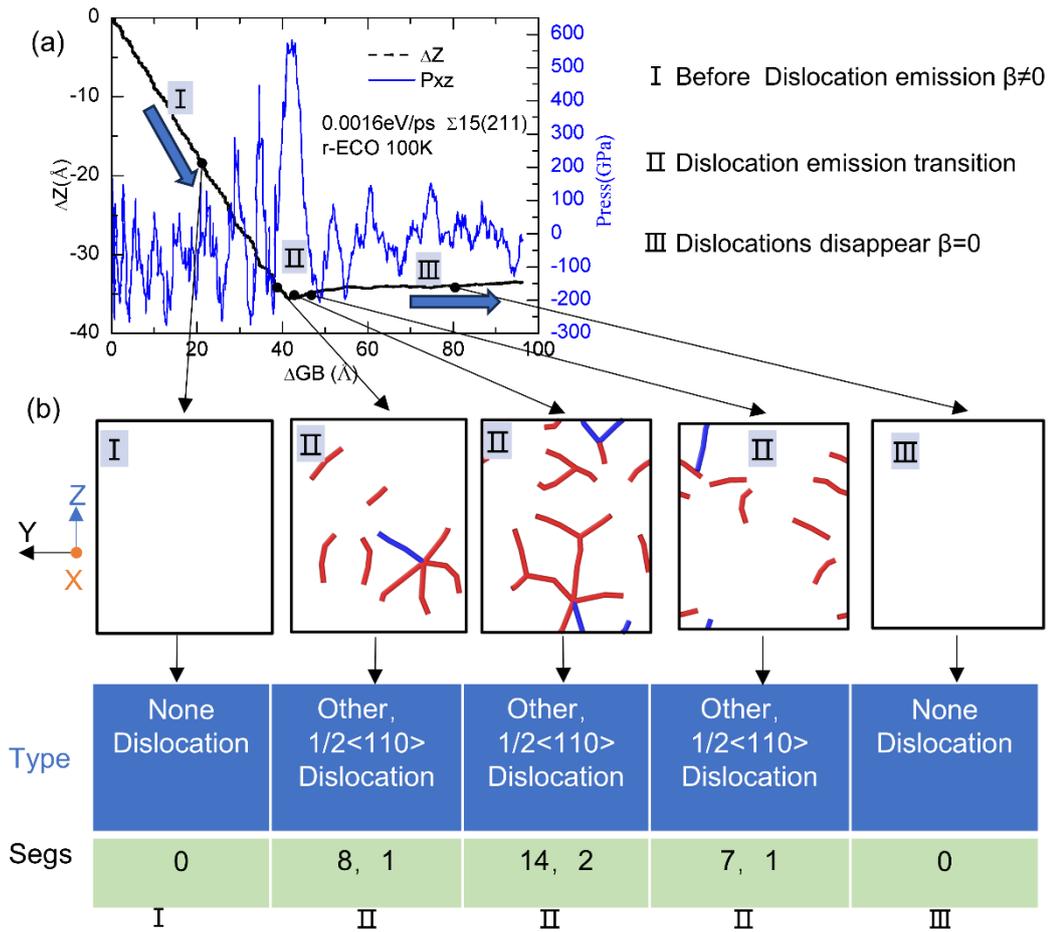

Fig.6. Complex shear coupling mode for Σ15(211) GB at a rate of 0.0016 eV/ps.

(a)~~GB translation in z direction~~ The lateral displacement of GB and press in shear coupling direction Pxz as a function of normal displacement of GB. (b) Evolution of dislocations at transition points for Σ15(211) GB. Dislocations are extracted by Dislocation Extraction Algorithm(DXA) in Ovito.Atoms are not shown.

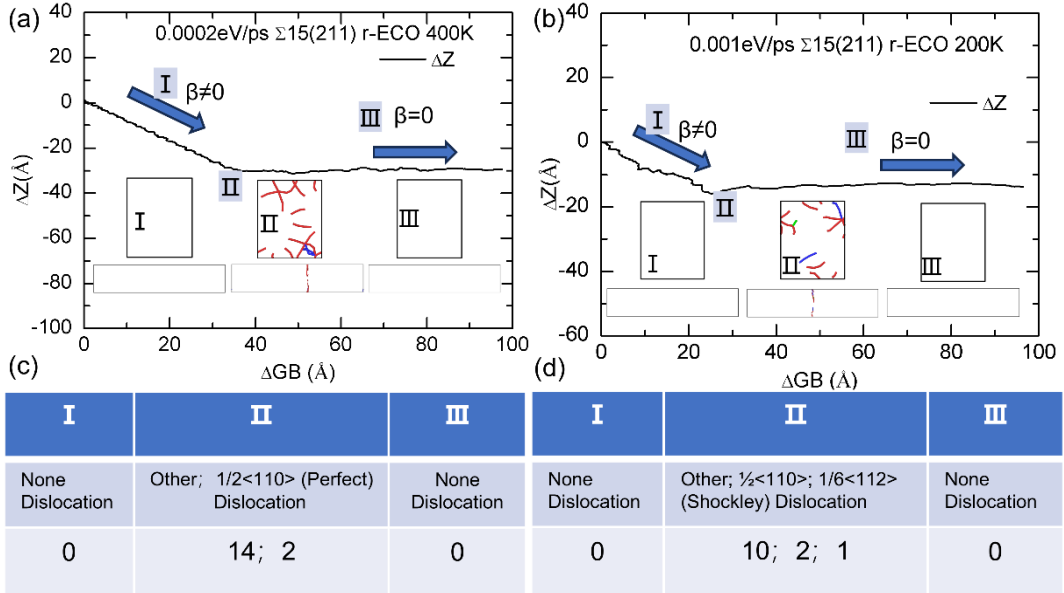

Fig.7. Complex shear coupling mode for Σ15(211) GB at rate of (a) 0.0002 eV/ps and (b) 0.001 eV/ps. Dislocations at the transition point of shear coupling pattern (c) 0.0002 eV/ps and (d) 0.001 eV/ps.

It is common to observe complex grain boundary (GB) migrations when there are changes in driving force or temperature. However, there are some similar migration behaviors for GBs, such as the transition from couple to only normal migration, which can be observed in Σ15(211) at various rates. To investigate the mechanisms behind this transition, the dislocations during GB migration were examined using the Dislocation Extraction Algorithm (DXA) in Ovito [19]. Interestingly, it was found that GBs transition from couple to only normal migration in the presence of dislocations nucleated within the GB at certain rates. These dislocations weren't observed during GB migration, except during the transition process from shear coupling to only normal migration. Once the transition is complete, these dislocations disappear again. Regarding the dislocation nucleation-assisted transition, shear-coupled GB migration is known as being mediated by the motion of GB disconnections with both step heights and non-zero Burgers vectors, while the only normal migration (or pure GB migration) is believed to be mediated by disconnections that have non-zero step heights but no Burgers vectors. Therefore, the nucleation of dislocations in GBs should be the byproduct of the transition of disconnections having Burgers vectors into disconnections without Burgers vectors. The dislocations compensate for the differences in Burgers vectors of these two types of disconnections.

Fig. 6 and Fig. 7 illustrate the presence of dislocations during GB migration at specific rates and temperatures. For example, at a rate of 0.0016 eV/ps at 100K and a rate of 0.0002 eV/ps at 400K, these dislocations include 1/2<110> perfect dislocations (blue lines) and other undistinguished dislocations (red lines). At a rate of 0.001 eV/ps at 200K, in addition to other dislocation styles and 1/2<110> perfect dislocations, there are 1/6<112> Shockley dislocations (green lines) present. It should be noted that Shockley dislocations can't climb at low temperatures, which may result in no translation in the z-direction. These dislocations influence the GB migration characteristics by reducing the translation in the z-direction. When the stress becomes too large, the GB needs to find a new relaxation mechanism, such as a change of shear coupling mode. Additionally, the release of dislocations can decrease stress and facilitate GB relaxation [20,21]. The emission of dislocations occurs at different rates, indicating that dislocation emission is a common method for GBs to relax stress and a means to achieve only normal migration.

## 4. Conclusions

In conclusion, the rate is the decisive factor for normal velocity, with higher rates leading to a higher normal velocity for $\Sigma3(110)$, $\Sigma15(211)$, and $\Sigma11(311)$ GBs. The inverse direction of shear coupling for $\Sigma15(211)$ occurred at a specific rate. With increasing rate, the only normal migration of the GB and the movement of the transition point were found and a linear relationship between the rate and the transition point during shear coupling migration was discovered. I have observed that a GB could migrate only normally in the presence of dislocations, which leads to depression of translation in the z direction and the relaxation of high stress for GBs. GB can dynamically tune its deformability via self-driven dislocation emission, which is strongly correlated with its stress state. These findings contribute to a deeper understanding of microstructural evolution and have implications for the design of materials with enhanced properties.

**Acknowledgments**

This research was supported by the Natural Science Foundation of the Jiangsu Higher Education Institutions of China (Grant No. 22KJD140001 and Grant No. 22KJD430002) and Joint Project of Industry-University-Research of Jiangsu Province (Grant No. BY20221296), This research was enabled by the use of computing resources provided by WestGrid and Compute/Calcul Canada.

**Author contributions**

**Tingting Yu**: conception, experimental design, carrying out measurements and manuscript composition.

**Conflicts of interest or competing interests**

There is no relationships or interests of the manuscript's authors that could potentially influence or bias the submitted work.

**Data and code availability**

Data available on request from the author.

**Supplementary information**

No supplementary information.

**Ethical approval**

There is no experiments involving human tissue.

**Statement**

During the preparation of this work, I used ChatGPT to improve readability and language. After using this tool, I reviewed and edited the content as needed and took full responsibility for the content of the publication.